\documentclass{article}
\usepackage[keeplastbox]{flushend}
\usepackage[preprint]{spconf}

\copyrightnotice{\begin{minipage}{\textwidth}\copyright\ 2021 IEEE. Personal use of this material is permitted. Permission from IEEE must be obtained for all other uses, in any current or future media, including reprinting/republishing this material for advertising or promotional purposes, creating new collective works, for resale or redistribution to servers or lists, or reuse of any copyrighted component of this work in other works.\end{minipage}}

\usepackage{amsmath,graphicx}  
\usepackage{lipsum}  
\usepackage{tikz}
\usepackage{amsthm,amssymb,amsmath,mathrsfs}
\usetikzlibrary{automata,arrows,positioning}
\usetikzlibrary{angles,quotes,calc}

  \newtheorem{thm}{\bf Theorem}

  \usepackage{dirtytalk}
\usepackage[]{amsmath}
\usepackage{amsfonts}
\usepackage{amssymb}
\usepackage{bbm}
\usepackage{breqn}
\usepackage{subcaption}
\usepackage{diagbox}
\usepackage{cite}
\usepackage{cases}
\usepackage{tabu}
\usepackage{url}
\usepackage{multicol}
\usepackage{color}
\usepackage[bottom]{footmisc}
\usepackage{environ}
\usepackage{tikz} 
\usepackage{enumerate}
\usepackage[]{algpseudocode}
\algtext*{EndWhile}
\algtext*{EndIf}
\algtext*{EndFor}
\usepackage{algorithm}

\ninept
\title{Physical-Layer Security via Distributed Beamforming in the Presence of Adversaries with Unknown Locations}

\name{ Yagiz Savas$^\dagger$,  Abolfazl Hashemi$^\dagger$,  Abraham P. Vinod$^\ddagger$, Brian M. Sadler$^\star$, and Ufuk Topcu$^\dagger$
\thanks{This work is supported by the collaborative agreement ARL DCIST CRA W911NF-17-2-0181.}}
\address{$^\dagger$ University of Texas at Austin, TX, USA\\
$^\ddagger$Mitsubishi Electric Research Laboratories, MA, USA\\
$^\star$U.S. Army Research Laboratory, MD, USA}

\date{}
\begin{document}
\maketitle
\begin{abstract}
We study the problem of securely communicating a sequence of information bits with a client in the presence of multiple adversaries at unknown locations in the environment. We assume that the client and the adversaries are located in the far-field region, and all possible directions for each adversary can be expressed as a continuous interval of directions. In such a setting, we develop a periodic transmission strategy, i.e., a sequence of joint beamforming gain and artificial noise pairs, that prevents the adversaries from decreasing their uncertainty on the information sequence by eavesdropping on the transmission. We formulate a series of nonconvex semi-infinite optimization problems to synthesize the transmission strategy. We show that the semi-definite program (SDP) relaxations of these nonconvex problems are exact under an efficiently verifiable sufficient condition. We approximate the SDP relaxations, which are subject to infinitely many constraints, by randomly sampling a finite subset of the constraints and establish the probability with which optimal solutions to the obtained finite SDPs and the semi-infinite SDPs coincide. We demonstrate with numerical simulations that the proposed periodic strategy can ensure the security of communication in scenarios in which all stationary strategies fail to guarantee security.
\end{abstract}
\section{Introduction}
We consider a group of agents, e.g., mobile robots, each of which carries an antenna. The agents are distributed in an environment from which they collect confidential data, i.e., a sequence of information bits. The agents' goal is to communicate the collected data wirelessly with a client, i.e., intended receiver, located in the far-field region. A known number of adversaries, i.e., unintended receivers, located also in the far-field region, eavesdrop on the transmission to infer the information bits. The exact directions of the adversaries are unknown to the agents, but we assume that all possible directions for each adversary can be conservatively expressed as a continuous interval of directions. In such a setting, we develop a sequential transmission strategy that enables the agents to communicate the data with the client while ensuring that no adversary can decrease its uncertainty on the information bits by eavesdropping.

Physical-layer security is an information-theoretic concept that enables the secure exchange of confidential data without relying on higher-level encryption \cite{mukherjee2014principles,shiu2011physical}. By exploiting the channel characteristics and designing suitable encoder-decoder pairs, agents can securely communicate with a client in the presence of adversaries \cite{wyner1975wire,ozarow1984wire,barros2006secrecy,csiszar1978broadcast,el2007wiretap}. In particular, it is shown in \cite{ozarow1984wire} that a sequence of $K$ information bits can be communicated securely in two steps as follows. First, the agents map the information bits into $N$ encoded symbols where $N$$\geq$$K$. Second, the encoded symbols are transmitted such that the client receives all the transmitted symbols, and no adversary receives more than $N$$-$$K$ symbols. The properties of the coding scheme can then be used to show that the uncertainty of the adversaries on the information bits remains unchanged after eavesdropping \cite{el2007wiretap,ozarow1984wire}. In this paper, we design a distributed beamforming strategy to transmit the encoded symbols such that the adversaries are guaranteed to receive no more than $N$$-$$K$ symbols despite their unknown directions.

Distributed beamforming (DB) is a wireless communication technique in which a group of agents collectively transmit a common signal \cite{Ochiai2005,visotsky1999optimum,mudumbai2009distributed}. DB traditionally concerns designing a vector of complex beamforming gains to adjust the phases of the transmitted signals so that the signal-to-interference-plus-noise ratio (SINR) received by the client is maximized \cite{mudumbai2009distributed2,ding2008distributed}. To prevent adversaries whose directions are unknown to the agents from receiving the signal, the agents may transmit the signal by jointly broadcasting an artificial noise (AN) to decrease the SINR at the adversaries \cite{goel2008guaranteeing,Justin}. When the agents know the adversary directions exactly, they can concentrate the AN specifically in those directions \cite{liao2010qos}. Moreover, when adversary directions are expressed as uncertainty sets, robust beamforming approaches can be used to keep the SINR at adversaries below a desired level \cite{gershman2010convex,lorenz2005robust}.

To the best of our knowledge, all DB designs in the literature yield stationary transmission strategies. That is, the agents transmit all encoded symbols using the same joint beamforming gain and AN pair. For secure communication, such strategies should satisfy the condition that each adversary in the environment receives zero symbol. Otherwise, an adversary that receives a single symbol can receive all the symbols due to the stationary nature of transmission. When the agents have limited transmission power, achieving such a restrictive condition may not be possible since it requires the SINR at all adversary directions to be low at the same time.

In this paper, we introduce a periodic transmission strategy, i.e., a sequence of joint beamforming gain and AN pairs, that allows the adversaries to receive some of the transmitted symbols but ensures that no adversary can receive more than $N$$-$$K$ symbols. When combined with a suitable encoder-decoder pair, such a strategy can make secure communication possible in scenarios in which the security cannot be ensured by stationary strategies. We show that one can synthesize the proposed strategy by solving a series of nonconvex semi-infinite optimization problems. We prove that the semi-definite program (SDP) relaxations \cite{gershman2010convex} of these nonconvex problems are exact under a sufficient condition which can be verified efficiently. Although the resulting SDPs are convex, they are subject to infinitely many constraints. By sampling a finite subset of these constraints, we construct finite SDPs. Finally, by using results from robust convex programming~\cite{campi2009scenario}, we establish the probability with which optimal solutions to the finite SDPs and the semi-infinite SDPs coincide.

\noindent\textbf{Notation:} We denote the size of a finite set $\mathcal{X}$ by $\lvert \mathcal{X} \rvert$ and the magnitude of a complex number $z$ by $|z|$. For $N$$\in$$\mathbb{N}$, $[N]$$=$$\{1,2,\ldots,N\}$. 
\section{Problem Formulation}
In this section, we introduce the considered environment and the communication model, and provide the formal problem statement.

\noindent\textbf{Environment:} We consider a group of $m$$\in$$\mathbb{N}$ agents who aim to communicate a common information block $S^K$$=$$(S_1,S_2,\ldots, S_K)$, where $K$$\in$$\mathbb{N}$, with a client in the presence of $L$$\in$$\mathbb{N}$ adversaries $\{a_i$$:$$i$$\in$$[L]\}$ in the environment. We make the standard assumption \cite{wyner1975wire,ozarow1984wire} that $S^K$ is a sequence of independent and identically distributed binary random variables with uniform distribution. The client is located in the \textit{known} far-field direction $\theta_c$$\in$$[-\pi,\pi)$. The exact directions of the adversaries located in the far-field region are \textit{unknown}; however, we assume that each adversary $a_i$ is associated with a continuous interval $I_i$$\subseteq$$[-\pi,\pi)$ representing all possible directions of $a_i$. Finally, we assume that there is an adversary-free direction interval $I_0$$=$$[\theta_c$$-$$\delta,\theta_c$$+$$\delta]$ where $\delta$$\in$$[-\pi,\pi)$ in the environment. Clearly, if an adversary is permitted in the interval $I_0$, then it receives the same message with the client, which makes secure communication with the client impossible.

\noindent \textbf{Encoder-decoder pair:} The agents map $S^K$ into an encoded block $X^N$$=$$(X_1,X_2,\ldots,X_N)$ of symbols, where $N$$\in$$\mathbb{N}$ such that $N$$\geq$$K$, using an encoder-decoder pair that is publicly known. We consider a coding scheme that is based on an $[N,N$$-$$K]$ linear maximum-distance-separable (MDS) code $\mathcal{C}$$\subseteq$$\mathbb{F}_q^{N}$ where $q$ is a large enough prime power \cite{ozarow1984wire}. We briefly discuss the relevant properties of the coding scheme and refer the reader to \cite{ozarow1984wire,el2007wiretap} for further details.


The coding scheme guarantees that if the client receives $X^N$, it can recover $S^K$ with probability one. Moreover, the knowledge of $\mu$$\leq$$N$$-$$K$ symbols of $X^N$ leaves the uncertainty of $S^K$ unchanged for the adversaries \cite{el2007wiretap}. In particular, consider an adversary $a_i$ that eavesdrops on the transmission of $X^N$. Suppose that the adversary observes the set  $\{X_t : t$$\in$$\mathcal{O}_i\}$ of symbols where $\mathcal{O}_i$$\subseteq$$[N]$ such that $\lvert \mathcal{O}_i\rvert$$=$$\mu_i$$\leq$$N$. The knowledge of the adversary on $X^N$ is described by the sequence $Z^{N,i}$$=$$(Z^i_1,Z^i_2,\ldots, Z^i_N)$ where 
\begin{align}\label{obs_seq}
    Z^i_t:=\begin{cases} X_t & \text{if}\  t\in \mathcal{O}_i\\
    \overline{X} & \text{if} \ t \not\in \mathcal{O}_i.
    \end{cases}
\end{align}
Here, $\overline{X}$ is a random variable with uniform distribution over $[q]$. The equivocation $\Delta_i$ measures the \textit{minimum} uncertainty of the adversary $a_i$ on $S^K$ after observing $\mu_i$ symbols of $X^N$ and is defined as \cite{ozarow1984wire}
\begin{align*}
    \Delta_i := \min_{\lvert \mathcal{O}_i \rvert = \mu_i}H( S^K | Z^{N,i}).
\end{align*}
$H(S^K | Z^{N,i})$ denotes the entropy of  $S^K$ conditioned on  $Z^{N,i}$ \cite{cover1999elements}. Note that $\Delta_i$$\leq$$H(S^K | Z^{N,i})$$\leq$$H(S^K)$$=$$K$. The coding scheme ensures that if $\mu_i$$\leq$$N$$-$$K$, then $\Delta_i$$=$$K$. That is, an adversary $a_i$ \textit{cannot decrease its uncertainty} on $S^K$ by observing $\mu_i$$\leq$$N$$-$$K$ symbols of $X^N$. In this paper, we design a transmission strategy such that the client receives $X^N$ and $\mu_i$$\leq$$N$$-$$K$ for all $i$$\in$$[L]$.

\noindent \textbf{Transmission:} At discrete time step $t$$\in$$[N]$, each agent transmits $X_t$ as a continuous signal $s_t$ using an ideal isotropic antenna with maximum transmit power $P$. We assume that the client and each adversary carry a single antenna; the signal $s_t$ propagates in free space with no reflection or scattering; there is no mutual coupling between the agents' antennas; and the time discretization is coarse enough to ensure that inter-symbol interference is negligible. In addition to free space propagation, the assumptions may also hold at longer wavelengths even when propagating through a complex environment \cite{choi2017low}. We express the narrowband wireless channel between the agent $i$$\in$$[m]$ and a receiver in the direction $\theta$$\in$$[-\pi,\pi)$ by a \textit{known} complex scalar gain $h_i(\theta)$. Finally, we assume that the local oscillators of all agents are time- and frequency-synchronized. 

The agents transmit the signal 
$y_{\text{transmit}}[t]$$:=$${\bf{w}}_t s_t$$+$${\bf{v}}_t$ at time step $t$$\in$$[N]$
where ${\bf{w}}_t$$\in$$\mathbb{C}^m$ is the vector of complex beamforming gains, and ${\bf{v}}_t$$\in$$\mathbb{C}^m$ is the vector of artificial noise, which follows the zero-mean complex Gaussian distribution with covariance matrix ${\bf \Sigma}_t$$\succcurlyeq$$0$. The pair $({\bf{w}}_t, {\bf \Sigma}_t)$ is the design variable that we choose to ensure the security of communication. Since each agent have a maximum transmit power $P$, assuming that $\mathbb{E}[s_t^2]$$=$$1$, we have $|\mathbf{w}_t(i)|^2$$+$${\bf \Sigma}_t(i,i)$$\leq$$P$ where $\mathbf{w}_t(i)$$\in$$\mathbb{C}$ and ${\bf \Sigma}_t(i,i)$$\in$$\mathbb{R}$ are the $i$-th component of $\mathbf{w}_t$ and $i$-th diagonal component of ${\bf \Sigma}_t$, respectively.

The transmitted signal travels through the channel, and the signal obtained by a receiver in the direction $\theta$$\in$$[-\pi,\pi)$ is given by 
\begin{align*}
    y^{\theta}_{\text{receive}}[t]&:={\bf{h}}(\theta)^{H} y_{\text{transmit}}[t]+ n_t
    \\ &= {\bf{h}}(\theta)^{H}{\bf{w}}_t s_t+ {\bf{h}}(\theta)^{H}{\bf{v}}_t+n_t
\end{align*}
where ${\bf{h}}(\theta)^H$$=$$[h^H_1(\theta),h^H_2(\theta),\ldots, h^H_m(\theta)]$, $n_t$ is additive noise with the distribution $\mathcal{N}(0,\sigma_n^2)$, and $(\cdot)^{H}$ denotes the cojugate transpose operation. Consequently, the signal-to-interference-plus-noise ratio (SINR) obtained by a receiver in the direction $\theta$$\in$$[-\pi,\pi)$ is 
\begin{align*}
    \text{SINR}_t(\theta)=\frac{{\bf{w}}_t^{H}{\bf{H}}(\theta){\bf{w}}_t}{\text{Tr}({\bf{H}}(\theta){\bf \Sigma}_t)+\sigma_n^2}
\end{align*}
where ${\bf{H}}(\theta)$$=$${\bf{h}}(\theta){\bf{h}}(\theta)^{H}$ and $\text{Tr}(\cdot)$ denotes the trace of a matrix.

Let $\gamma_c,\gamma_a$$\in$$\mathbb{R}$ be positive constants such that $\gamma_c$$>$$\gamma_a$. For simplicity, we assume that, if $\text{SINR}_t(\theta)$$\geq$$\gamma_c$, the receiver in the direction $\theta$ observes $X_t$ with probability one; and if $\text{SINR}_t(\theta)$$\leq$$\gamma_a$, the receiver observes $X_t$ with probability zero. In practice, a small probability of error for the client and a large probability of error for the adversaries can be achieved using carefully chosen $\gamma_c$ and $\gamma_a$ values.

\noindent\textbf{Problem Statement:} 
For each $i$$\in$$[L]$, let 
\begin{align*}
    \mathcal{O}_i=\{t\in[N]: \text{SINR}_t(\theta)>\gamma_a\ \  \text{for some}\ \  \theta\in I_i\}
\end{align*}
be the set of time steps that the adversary $a_i$ observes the transmitted symbol with a nonzero probability. Note that $a_i$'s \textit{maximum} knowledge on $X^N$ is described by the sequence $Z^{N,i}$ defined in \eqref{obs_seq}. The secure communication problem is then to design a sequential transmission strategy $T^{\star}$$=$$(({\bf{w}}^{\star}_1,{\bf \Sigma}^{\star}_1), ({\bf{w}}^{\star}_2,{\bf \Sigma}^{\star}_2), \ldots, ({\bf{w}}^{\star}_N,{\bf \Sigma}^{\star}_N))$ such that the client observes all transmitted symbols with probability one, i.e., $\min_{t\in [N]}\text{SINR}_t(\theta_c)\geq\gamma_c$, and $\lvert \mathcal{O}_i \rvert$$\leq$$N$$-$$K$ for all $i$$\in$$[L]$. 


\section{Secure Communication in the Presence of Adversaries}
In this section, we present a solution to the secure communication problem with certain probabilistic guarantees. We first consider an approach in which the agents transmit $S^K$ without encoding through a stationary transmission strategy. We argue that such an approach may, in general, yield infeasible solutions to the secure communication problem due to the agents' individual power constraints. Next, we show that one can encode $S^K$ into $X^N$ where $N$$=$$LK$ and obtain a periodic transmission strategy $T^{\star}$ that decreases the minimum power required for the transmission of each $X_t$. Finally, we show that a strategy $T^{\star}$ with probabilistic approximation guarantees can be synthesized by solving a series of semi-definite programs (SDPs).
\subsection{Transmission Without Encoding}
In practice, it is desirable that the agents encode $S^K$ into $X^N$ with maximum information rate $R$$=$$K/N$ so that the redundancy in the transmission and overall power consumption are minimized. Specifically, if the client receives each transmitted symbol with probability one, i.e., $\text{SINR}_t(\theta_c)$$\geq$$\gamma_c$, the agents can choose $R$$=$$1$ and transmit $S^K$ to the client without encoding. However, if $R$$=$$1$ in the presence of adversaries $\{a_i$$:$$i$$\in$$[L]\}$, to ensure that no adversary can decrease its uncertainty on $S^K$ by eavesdropping, i.e., $\min_{i\in [L]}\Delta_i$$=$$K$, the agents should follow a strategy $T^{\star}$ such that $\mu_i$$=$$0$ for all $i$$\in$$[L]$. Such a strategy that minimizes the overall power consumption can be synthesized by solving the problem 
\begin{subequations}
\begin{align}\label{nonencode_1}
\textbf{(P1)}:\ \ \ \min_{{\bf{w}}_t,{\bf \Sigma}_t\succcurlyeq 0}\ \ & \lVert{\bf{w}}_t\rVert^2+  \text{Tr}({\bf \Sigma}_t)\\
  \text{subject to:} \ \ & \text{SINR}_t(\theta_c)\geq \gamma_c\\
    &\text{SINR}_t(\theta)\leq \gamma_a \ \ \qquad \quad \  \forall \theta \in \cup_{i\in [L]} I_i \label{nonencode_cons} \\
&    |\mathbf{w}_t(i)|^2+{\bf \Sigma}_t(i,i)\leq P \ \ \ \forall i\in [m]. \label{nonencode_last}
\end{align}
\end{subequations}
Let $(\overline{{\bf{w}}},\overline{{\bf \Sigma}})$ be a solution to the above optimization problem. Then, the stationary strategy $T^{\star}$ where $({\bf{w}}^{\star}_t,{\bf \Sigma}^{\star}_t)$$=$$(\overline{{\bf{w}}},\overline{{\bf \Sigma}})$ for all $t$$\in$$[K]$ is a solution to the secure communication problem. 

Note that a feasible solution to \eqref{nonencode_1}-\eqref{nonencode_last} must satisfy the condition that $\text{SINR}_t(\theta)$$\leq$$\gamma_a$ \textit{for all possible directions $I_i$ for each adversary $i$$\in$$[L]$}. Due to the individual power constraints in \eqref{nonencode_last}, the set of feasible solutions to \eqref{nonencode_1}-\eqref{nonencode_last} can be empty in many cases. One approach to increase the size of the feasible set might be to increase the redundancy in the transmission by choosing $N$$>$$K$ and allowing the adversaries to observe some of the transmitted symbols as long as no adversary can decrease its uncertainty on $S^K$ using those observations. In the next section, we provide a method to increase the size of the feasible set while ensuring the security of communication.

\subsection{Transmission with Encoding}
Recall that the agents employ an $[N,N$$-$$K]$ linear MDS code to guarantee that no adversary $i$$\in$$[L]$ can decrease its uncertainty on $S^K$ by observing at most $\mu_i$$\leq$$N$$-$$K$ symbols of $X^N$. A consequence of this coding scheme is that, by choosing $N$$>$$K$, the agents can let the adversaries to observe some of the transmitted symbols, i.e., $\mu_i$$>$$0$, while still ensuring that $\Delta_i$$=$$K$ for all $i$$\in$$[L]$. 


Suppose that we choose $N$$=$$LK$. Then, the strategy $T^{\star}$ must ensure that $\mu_i$$\leq$$(L$$-$$1)K$ for each $i$$\in$$[L]$. One can synthesize such a strategy $T^{\star}$ by focusing on the strategies that are periodic in $L$, i.e., $({\bf{w}}_{t_1}, {\bf \Sigma}_{t_1})$$=$$({\bf{w}}_{t_2}, {\bf \Sigma}_{t_2})$ if $t_1$$=$$(t_2 \mod L)$. Specifically, consider the following optimization problem where $\hat{t}$$:=$$(t$$\mod$$L)$.
\begin{subequations}
\begin{align}\label{encode_1}
\textbf{(P2)}:\ \ \ \min_{{\bf{w}}_t,{\bf \Sigma}_t\succcurlyeq 0}\ \ & \lVert{\bf{w}}_t\rVert^2+  \text{Tr}({\bf \Sigma}_t)\\ \label{encode_2}
  \text{subject to:} \ \ & \text{SINR}_t(\theta_c)\geq \gamma_c\\
    &\text{SINR}_t(\theta)\leq \gamma_a \ \ \qquad \quad \  \forall \theta \in I_{\hat{t}} \label{encode_cons} \\
&    |\mathbf{w}_t(i)|^2+{\bf \Sigma}_t(i,i)\leq P \ \ \ \forall i\in [m] \label{encode_last}
\end{align}
\end{subequations}
The constraint in \eqref{encode_cons} is the same for each $t_1,t_2$$\in$$[N]$ that satisfies $t_1$$=$$(t_2 \mod L)$, and all the other constraints are the same for all $t$$\in$$[N]$. Let $(\widetilde{{\bf{w}}}_t,\widetilde{{\bf \Sigma}}_t)$ be an optimal solution to \textbf{P2}. The \textit{periodic} strategy $T^{\star}$ where $({\bf{w}}^{\star}_t,{\bf \Sigma}^{\star}_t)$$=$$(\widetilde{{\bf{w}}}_t,\widetilde{{\bf \Sigma}}_t)$ ensures that each adversary observes at most $(L$$-$$1)$ of $L$ transmitted symbols with a nonzero probability. Hence, $T^{\star}$ constitutes a solution to the secure communication problem as we have $\max_{i\in[L]}\mu_i$$\leq$$(L-1)K$ for all $i$$\in$$[L]$.

There are two important properties of the communication strategy described above. First, the information rate of transmission is $R$$=$$1/L$. Therefore, if there are multiple adversaries in the environment, the agents strategically introduce redundancy to the transmission so that no adversary can decrease its uncertainty on the information block through eavesdropping. Second, we have \textbf{P2}$^{\star}$$\leq$$\textbf{P1}^{\star}$ where \textbf{P1}$^{\star}$ and \textbf{P2}$^{\star}$ are the optimal values of the problems \textbf{P1} and \textbf{P2}, respectively. The inequality follows from the fact that any feasible solution to \textbf{P1} is also a feasible solution to \textbf{P2}. Hence, \textit{by combining encoding with a periodic transmission strategy, we may establish a secure communication link with the client, which would not be possible to achieve using stationary transmission strategies.}

\subsection{SDP Relaxation and Probabilistic Approximation}
In the previous sections, we introduced two secure transmission strategies whose synthesis requires one to solve the optimization problems \textbf{P1} and \textbf{P2} which are nonconvex optimization problems with infinitely many constraints. In this section, we present a method to solve \textbf{P2} efficiently with probabilistic approximation guarantees. The same method can be used to solve \textbf{P1} with minor modifications.

\begin{figure*}[t]
	\centering
	\minipage[t]{1\linewidth}
	\begin{subfigure}[t]{.32\linewidth}
		\includegraphics[width=\textwidth]{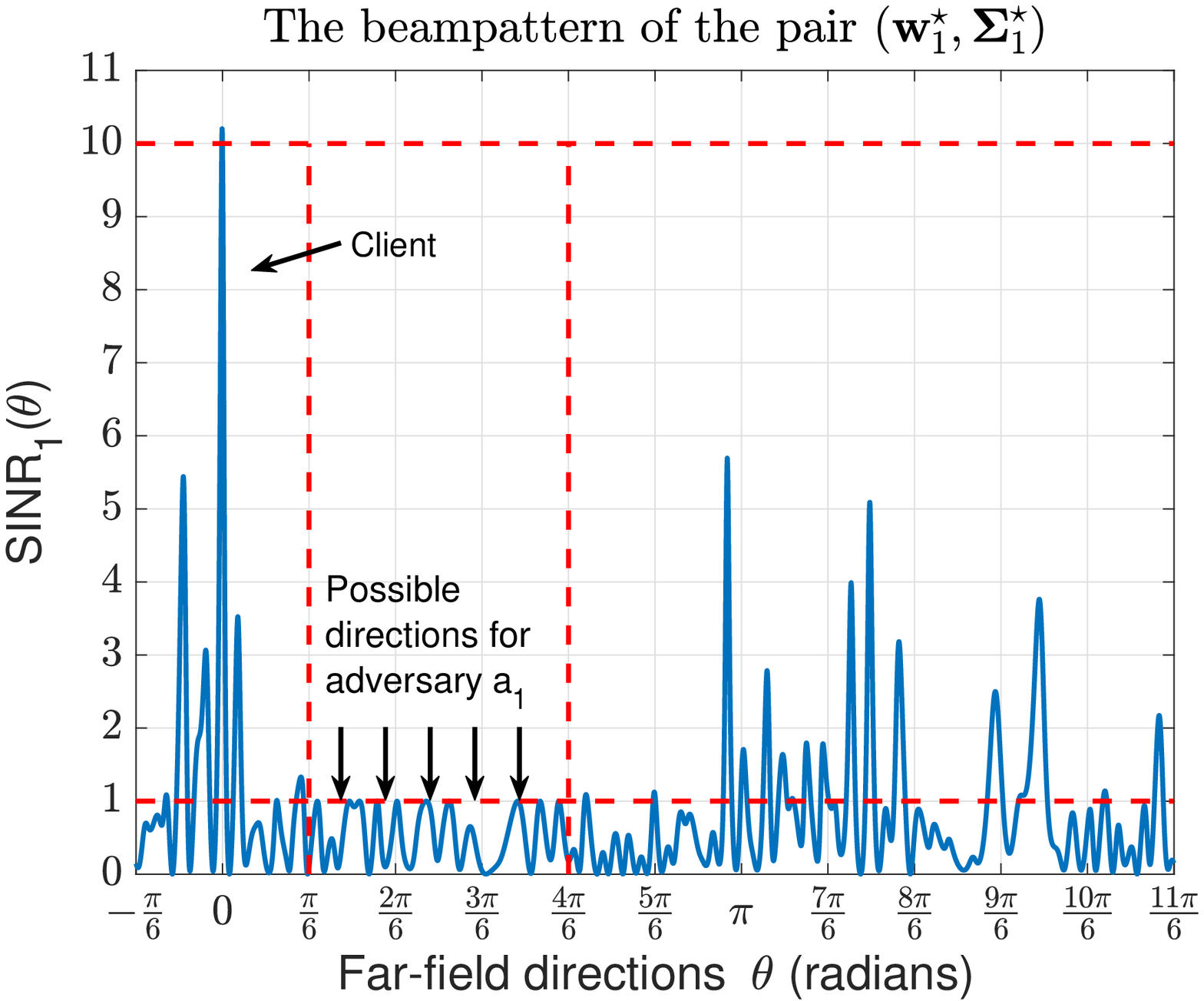}
	\end{subfigure}
	\begin{subfigure}[t]{.32\linewidth}
		\includegraphics[width=\linewidth]{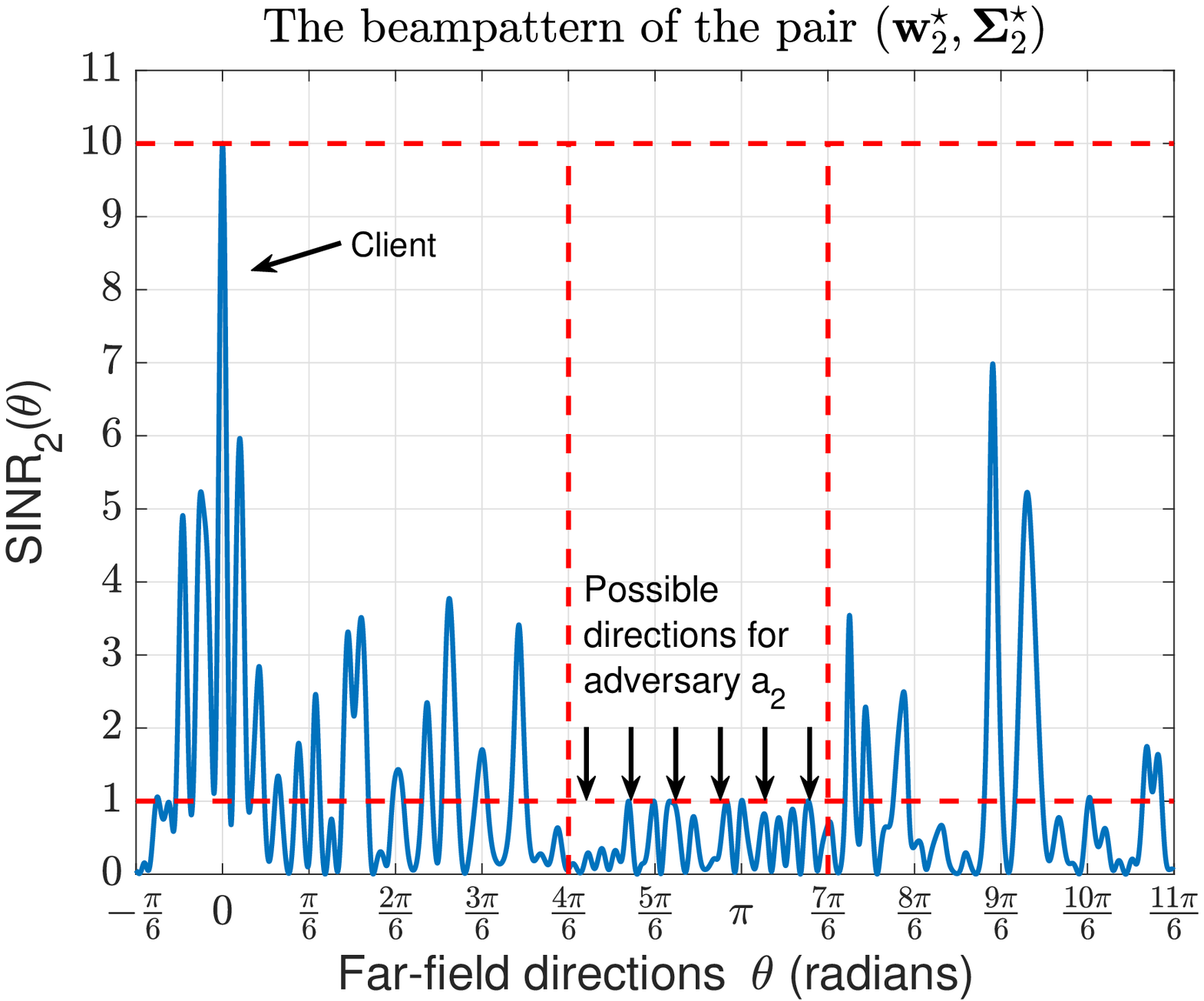}
	\end{subfigure}
	\begin{subfigure}[t]{.32\linewidth}
		\includegraphics[width=\linewidth]{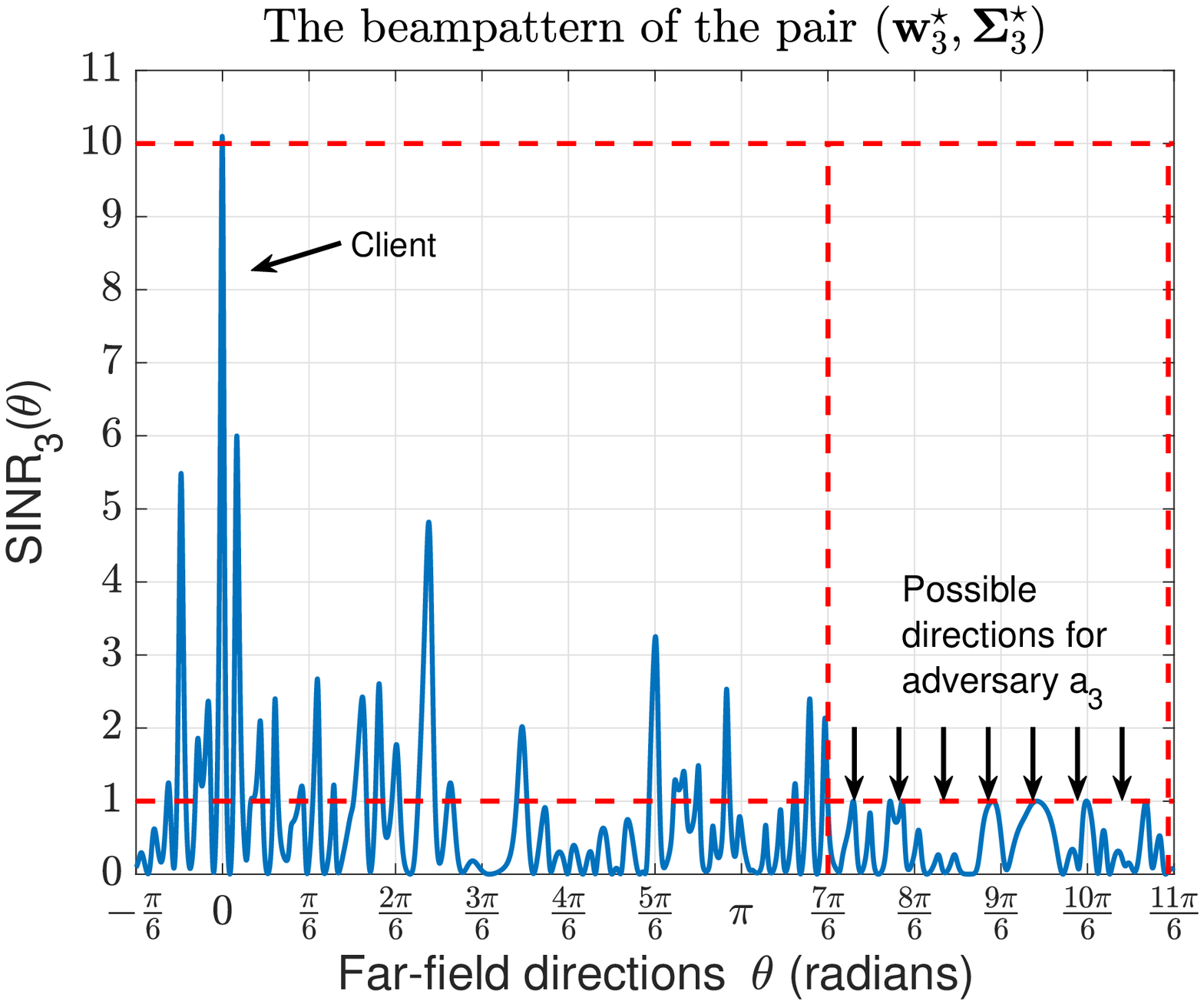}
	\end{subfigure}
	\caption{Beampatterns of the proposed periodic transmission strategy. The agents map $K$ information bits into $N$$=$$3K$ encoded symbols and transmit the symbols through the periodic strategy $T^{\star}$$=$$(({\bf{w}}^{\star}_1,{\bf \Sigma}^{\star}_1),({\bf{w}}^{\star}_2,{\bf \Sigma}^{\star}_2),({\bf{w}}^{\star}_3,{\bf \Sigma}^{\star}_3),({\bf{w}}^{\star}_1,{\bf \Sigma}^{\star}_1),({\bf{w}}^{\star}_2,{\bf \Sigma}^{\star}_2),\ldots)$. Note that the client observes all transmitted symbols since $\text{SINR}_t(\theta_c)$$\geq$$10$ for all $t$$\in$$[N]$. Moreover, no adversary observes more than 2/3 of the transmitted symbols since, for each $t$$\in$$\{1,2,3\}$, $\text{SINR}_t(\theta)$$\leq$$1$ for all $\theta$$\in$$I_t$.	}
	\label{fig:1}
	\endminipage 
	\vspace{-0.08cm}
\end{figure*}



We consider the SDP relaxation of \textbf{P2}, which is given as follows:
\begin{subequations}
\begin{align}\label{P2_obj}
       {(\textbf{P3}):} \min_{{\bf{W}}_t,{\bf \Sigma}_t\succcurlyeq 0} &\  \text{Tr}({\bf{W}}_t)+  \text{Tr}({\bf \Sigma}_t)\\
   \text{subject to :}&\ \label{P2_c1} \text{Tr}({\bf{H}}(\theta_c){\bf{W}}_t) \geq \gamma_c\text{Tr}({\bf{H}}(\theta_c){\bf \Sigma}_t)+\gamma_c\sigma_n^2 \\  \label{P2_c2}
  \forall \theta\in I_{\hat{t}}, \ \  &\ \text{Tr}({\bf{H}}(\theta){\bf{W}}_t) \leq \gamma_a\text{Tr}({\bf{H}}(\theta){\bf \Sigma}_t)+\gamma_a\sigma_n^2\\ \label{P2_c3}
 \forall i\in [m], \ \ & \ {\bf{W}}_t(i,i)+{\bf \Sigma}_t(i,i)\leq P.
\end{align}
\end{subequations}

Since \textbf{P3} is a relaxation of \textbf{P2}, we have $\textbf{P3}^{\star}$$\leq$$\textbf{P2}^{\star}$ where $\textbf{P3}^{\star}$ is the optimal value of $\textbf{P3}$. The following result establishes that, under mild assumptions, \textbf{P3} is an exact relaxation of \textbf{P2}, i.e., $\textbf{P2}^{\star}$$=$$\textbf{P3}^{\star}$.
{\setlength{\parindent}{0cm}\noindent 
\begin{thm}\label{rank_one_thm}
Suppose that \textbf{P3} has a feasible solution $(\underline{{\bf{W}}}_t,\underline{{\bf \Sigma}}_t)$ such that $\text{rank}(\underline{{\bf{W}}}_t)$$=$$\text{rank}(\underline{\bf \Sigma}_t)$$=$$m$, and let $(\widetilde{{\bf{W}}}_t,\widetilde{{\bf \Sigma}}_t)$ be an optimal solution to \textbf{P3}. Then, we have $\text{rank}(\widetilde{{\bf{W}}}_t)$$=$$1$. Furthermore, if $\widetilde{{\bf{W}}}_t$$=$$\widetilde{{\bf{w}}}_t(\widetilde{{\bf{w}}}_t)^H$, then $(\widetilde{{\bf{w}}}_t,\widetilde{{\bf \Sigma}}_t)$ is an optimal solution to \textbf{P2}.
\end{thm}}
\noindent \textbf{Proof Sketch:} If \textbf{P3} has a feasible solution $(\underline{{\bf{W}}}_t,\underline{{\bf \Sigma}}_t)$ such that $\text{rank}(\underline{{\bf{W}}}_t)$$=$$\text{rank}(\underline{\bf \Sigma}_t)$$=$$m$, it can be shown that the Slater's condition \cite{lopez2007semi} holds for \textbf{P3}. Then, Karush-Kuhn-Tucker (KKT) conditions are necessary for the optimality of a point in \textbf{P3}. Let ${\bf{Y}}^{\star}$$\succcurlyeq$$0$, $\lambda^{\star}$$\geq$$0$, and the diagonal matrix ${\bf{Q}}^{\star}$$\succcurlyeq$$0$ be the optimal dual variables associated with the constraints ${\bf{W}}_t$$\succcurlyeq$$0$, \eqref{P2_c1}, and \eqref{P2_c3}, respectively. KKT conditions imply that there exist optimal dual variables $\mu^{\star}_1,\mu^{\star}_2,\ldots,\mu_k^{\star}$$\geq$$0$, where $k$$\leq$$2m^2$, that are associated with the directions $\theta_1,\theta_2,\ldots,\theta_k$$\in$$I_{\hat{t}}$ in \eqref{P2_c2} and satisfy the condition ${\bf{Y}}^{\star}=\mathbb{I}_{m\times m}+{\bf{Q}}^{\star} +\sum_{j=1}^k\mu^{\star}_j{\bf{H}}(\theta_j)-\lambda^{\star}{\bf{H}}(\theta_c)\succcurlyeq 0$. The result then follows from the proof of Proposition 1 in \cite{liao2010qos}. $\blacksquare$

For a given \textbf{P3} instance, correctness of the full rank condition in Theorem \ref{rank_one_thm} can be verified efficiently as follows. First, we introduce three new variables $v,v_1,v_2$$\in$$\mathbb{R}$ to \textbf{P3} such that ${\bf{W}}_t$$\succcurlyeq$$v_1 \mathbb{I}_{m\times m}$, ${{\bf \Sigma}}_t$$\succcurlyeq$$v_2 \mathbb{I}_{m \times m}$, and $v$$\leq$$\min \{v_1,v_2\}$. Next, we replace the objective in \eqref{P2_obj} with $\max_{{\bf{W}_t},{\bf \Sigma}_t \succcurlyeq 0} v$. The rank condition holds if and only if the optimal value of this program satisfies $v^{\star}$$>$$0$. In our extensive numerical experiments, we observe that when \textbf{P3} is feasible, there always exists a feasible solution satisfying the full rank condition. 

 An optimal solution to \textbf{P2} can be obtained from a solution of \textbf{P3} via the construction $\widetilde{{\bf{W}}}_t$$=$$\widetilde{{\bf{w}}}_t(\widetilde{{\bf{w}}}_t)^H$. However, since \textbf{P3} includes infinitely many constraints in \eqref{P2_c2}, in general, we cannot solve it exactly. We approximate \textbf{P3} by sampling $B$$\in$$\mathbb{N}$ points from the set $I_{\hat{t}}$ uniformly randomly. Let $\Theta_{B,\hat{t}}$$=$$\{\theta_i$$\in$$ I_{\hat{t}}: i$$\in$$B\}$. We define the approximate problem as follows: 
\begin{subequations}
\begin{align}\label{SDP_start}
       {(\textbf{P4}):} \min_{{\bf{W}}_t,{\bf \Sigma}_t\succcurlyeq 0} &\  \text{Tr}({\bf{W}}_t)+  \text{Tr}({\bf \Sigma}_t)\\
   \text{subject to :}&\ \label{P3_c1} \text{Tr}({\bf{H}}(\theta_c){\bf{W}}_t) \geq \gamma_c\text{Tr}({\bf{H}}(\theta_c){\bf \Sigma}_t)+\gamma_c\sigma_n^2 \\  \label{P3_c2}
 \forall \theta_i\in \Theta_{B,\hat{t}}, \ \  &\ \text{Tr}({\bf{H}}(\theta_i){\bf{W}}_t) \leq \gamma_a\text{Tr}({\bf{H}}(\theta_i){\bf \Sigma}_t)+\gamma_a\sigma_n^2\\ \label{SDP_end}
  \forall i\in [m], \ \ & \ {\bf{W}}_t(i,i)+{\bf \Sigma}_t(i,i)\leq P.
\end{align}
\end{subequations}
The problem \textbf{P4} can be solved efficiently through off-the-shelf solvers such as Gurobi \cite{gurobi} and SeDuMi \cite{sedumi}. Since \textbf{P4} includes only a subset of the constraints in \textbf{P3}, we have $\textbf{P4}^{\star}$$\leq$$\textbf{P3}^{\star}$ where $\textbf{P4}^{\star}$ denotes the optimal value of \textbf{P4}. Moreover, Theorem \ref{rank_one_thm} implies that any optimal solution $(\hat{{\bf{W}}}_t,\hat{{\bf \Sigma}}_t)$ to \textbf{P4} satisfies that $\text{rank}(\hat{{\bf{W}}}_t)$$=$$1$.

Finally, we need to establish the relation between the problems \textbf{P3} and \textbf{P4} to characterize the performance of a periodic transmission strategy synthesized from the solution of \textbf{P4}. Let $f$$:$$I_{\hat{t}}$$\rightarrow$$\mathbb{R}$ be a function such that $f(\theta):=\text{Tr}({\bf{H}}(\theta)\hat{{\bf{W}}}_t)- \gamma_a\text{Tr}({\bf{H}}(\theta)\hat{{{\bf \Sigma}}}_t)-\gamma_a\sigma_n^2$.
If $f(\theta)$$\leq$$0$ for all $\theta$$\in$$I_{\hat{t}}$, then we have $\textbf{P4}^{\star}$$=$$\textbf{P2}^{\star}$, and a solution to \textbf{P2} can be obtained from $(\hat{{\bf{W}}}_t,\hat{{\bf \Sigma}}_t)$ through the construction explained in Theorem \ref{rank_one_thm}. However, in general, it is possible that $f(\theta)$$>$$0$ for some $\theta$$\in$$I_{\hat{t}}$, in which case a feasible solution to \textbf{P2} cannot be obtained from $(\hat{{\bf{W}}}_t,\hat{{\bf \Sigma}}_t)$. Next, we characterize the probability of such an event in terms of the sample size $B$. Let $\text{Pr}(\{\theta$$\in$$I_{\hat{t}} : f(\theta)$$\geq$$0\})$ be the probability that the solution of \textbf{P4} violates the constraint in \eqref{P2_c2}.

{\setlength{\parindent}{0cm}\noindent 
\begin{thm}\label{sampling_thm} \cite{campi2009scenario}
Given a violation parameter $\beta_1$$\in$$(0,1)$ and a confidence parameter $\beta_2$$\in$$(0,1)$, if $B$$\geq$$(2\log_e(\beta_2^{-1})$$+$$16m^2)/\beta_1$,
then $\text{Pr}(\{\theta$$\in$$I_{\hat{t}}$$:$$f(\theta)$$\geq$$0\})$$\leq$$\beta_1$ with probability at least $1$$-$$\beta_2$.
\end{thm}}

The above result is a special case of Theorem 1 in \cite{campi2009scenario} which establishes the probability with which a solution to the finite approximation of a semi-infinite convex optimization problem is feasible for the original problem. Theorem \ref{sampling_thm} implies that, the following statement is true with probability at least $1$$-$$\beta_2$: if the number of samples $B$ from the set $I_{\hat{t}}$ satisfies the inequality in Theorem \ref{sampling_thm}, then a transmission strategy $T^{\star}$ can be synthesized from the solution of \textbf{P4} with probability at least $1$$-$$\beta_1$. We note that the lower bound on the sample size $B$ to ensure the probabilistic guarantee $1$$-$$\beta_1$ can be considerably large, e.g., $B$$\geq$$10^5$ for $\beta_1$$=$$\beta_2$$=$$0.01$ and $m$$=$$10$. However, we observe in our numerical experiments that, it is, in general, sufficient to choose $B$ in the order of $10^3$ to determine whether an optimal solution to \textbf{P4} is feasible for \textbf{P3}.

\section{Numerical Simulations}
We demonstrate the performance of the proposed strategies on a scenario in which $m$$=$$6$ agents aim to securely communicate with a client in the presence of $L$$=$$3$ adversaries. The agents are located randomly within a disk of radius 80 meters and carry antennas with maximum transmit power $P$$=$$1$ that operate at the carrier frequency $f_c$$=$$40$ MHz. Both the adversaries and the client are located on a circle of radius 300 meters. The direction of the client is $\theta_c$$=$$0$, the adversary-free interval is $I_0$$=$$(-\pi/6,\pi/6)$, and the direction intervals of the adversaries $a_1$, $a_2$, and $a_3$ are $I_1$$=$$[\pi/6,4\pi/6]$, $I_2$$=$$[4\pi/6,7\pi/6]$, and $I_3$$=$$[7\pi/6,11\pi/6]$, respectively. Following \cite{Ochiai2005}, we represent the channel $h_i(\theta)$ between the agent $i$$\in$$[m]$ and a receiver located in the direction $\theta$$\in$$[-\pi,\pi)$ with $h_i(\theta)$$=$$\exp(-j 2\pi f_c d_i(\theta) )$ where $d_i(\theta)$ is the Euclidian distance between the agent and the receiver. We set $\gamma_c$$=$$10$, $\gamma_a$$=$$1$, and $\sigma_n^2$$=$$1$ in order to achieve an SINR difference of at least 10 dBs between the client and the adversaries. Finally, we set $B$$=$$1000$ to obtain the finite SDPs corresponding to the problems \textbf{P1} and \textbf{P2}.

We first synthesized a stationary transmission strategy by solving the finite SDP corresponding to the problem \textbf{P1}. We observed that this problem has no feasible solution. In other words, in the considered scenario, \textit{there is no stationary transmission strategy that can be used by the agents to securely communicate with the client}. We then synthesized a periodic transmission strategy by solving the problem in \eqref{SDP_start}-\eqref{SDP_end} for each $t$$=$$\{1,2,3\}$. The resulting beampatterns are illustrated in Figure \ref{fig:1}. As can be seen from the figure, \textit{although the exact locations of the adversaries are unknown, the periodic transmission strategy guarantees that each adversary observes at most 2/3 of the transmitted symbols}. Consequently, it follows from the properties of the MDS codes that the agents can securely communicate with the client using the synthesized transmission strategy. 
\section{Conclusions}
We showed that an agent group can securely communicate with a client in the presence of $L$ adversaries by mapping $K$ information bits into $N$$=$$LK$ encoded symbols via $[N,N$$-$$K]$ linear maximum-distance-separable codes and using a periodic transmission strategy. Assuming the availability of the perfect channel state information (CSI), we provided a semi-definite program-based approach to synthesize the transmission strategy with probabilistic guarantees. Future work will focus on developing strategies that ensure the security of communication when the perfect CSI is unavailable at the agents.
\clearpage
\newpage
\bibliographystyle{IEEEtran}
\bibliography{main.bib}

\begin{thebibliography}{10}
\providecommand{\url}[1]{#1}
\csname url@samestyle\endcsname
\providecommand{\newblock}{\relax}
\providecommand{\bibinfo}[2]{#2}
\providecommand{\BIBentrySTDinterwordspacing}{\spaceskip=0pt\relax}
\providecommand{\BIBentryALTinterwordstretchfactor}{4}
\providecommand{\BIBentryALTinterwordspacing}{\spaceskip=\fontdimen2\font plus
\BIBentryALTinterwordstretchfactor\fontdimen3\font minus
  \fontdimen4\font\relax}
\providecommand{\BIBforeignlanguage}[2]{{%
\expandafter\ifx\csname l@#1\endcsname\relax
\typeout{** WARNING: IEEEtran.bst: No hyphenation pattern has been}%
\typeout{** loaded for the language `#1'. Using the pattern for}%
\typeout{** the default language instead.}%
\else
\language=\csname l@#1\endcsname
\fi
#2}}
\providecommand{\BIBdecl}{\relax}
\BIBdecl

\bibitem{mukherjee2014principles}
A.~Mukherjee, S.~A.~A. Fakoorian, J.~Huang, and A.~L. Swindlehurst,
  ``Principles of physical-layer security in multi user wireless networks: A
  survey,'' \emph{IEEE Communications Surveys \& Tutorials}, vol.~16, no.~3,
  pp. 1550--1573, 2014.

\bibitem{shiu2011physical}
Y.-S. Shiu, S.~Y. Chang, H.-C. Wu, S.~C.-H. Huang, and H.-H. Chen,
  ``Physical-layer security in wireless networks: A tutorial,'' \emph{IEEE
  Wireless Communications}, vol.~18, no.~2, pp. 66--74, 2011.

\bibitem{wyner1975wire}
A.~D. Wyner, ``The wire-tap channel,'' \emph{Bell System Technical Journal},
  vol.~54, no.~8, pp. 1355--1387, 1975.

\bibitem{ozarow1984wire}
L.~H. Ozarow and A.~D. Wyner, ``Wire-tap channel {II},'' \emph{AT\&T Bell
  Laboratories Technical Journal}, vol.~63, no.~10, pp. 2135--2157, 1984.

\bibitem{barros2006secrecy}
J.~Barros and M.~R. Rodrigues, ``Secrecy capacity of wireless channels,'' in
  \emph{IEEE International Symposium on Information Theory}, 2006, pp.
  356--360.

\bibitem{csiszar1978broadcast}
I.~Csisz{\'a}r and J.~Korner, ``Broadcast channels with confidential
  messages,'' \emph{IEEE Transactions on Information Theory}, vol.~24, no.~3,
  pp. 339--348, 1978.

\bibitem{el2007wiretap}
S.~Y. El~Rouayheb and E.~Soljanin, ``On wiretap networks {II},'' in \emph{IEEE
  International Symposium on Information Theory}, 2007, pp. 551--555.

\bibitem{Ochiai2005}
H.~Ochiai, P.~Mitran, H.~V. Poor, and V.~Tarokh, ``Collaborative beamforming
  for distributed wireless ad hoc sensor networks,'' \emph{IEEE Transactions on
  Signal Processing}, 2005.

\bibitem{visotsky1999optimum}
E.~Visotsky and U.~Madhow, ``Optimum beamforming using transmit antenna
  arrays,'' in \emph{IEEE Vehicular Technology Conference}, vol.~1, 1999, pp.
  851--856.

\bibitem{mudumbai2009distributed}
R.~Mudumbai, D.~R.~B. Iii, U.~Madhow, and H.~V. Poor, ``Distributed transmit
  beamforming: challenges and recent progress,'' \emph{IEEE Communications
  Magazine}, vol.~47, no.~2, pp. 102--110, 2009.

\bibitem{mudumbai2009distributed2}
R.~Mudumbai, J.~Hespanha, U.~Madhow, and G.~Barriac, ``Distributed transmit
  beamforming using feedback control,'' \emph{IEEE Transactions on Information
  Theory}, vol.~56, no.~1, pp. 411--426, 2009.

\bibitem{ding2008distributed}
Z.~Ding, W.~H. Chin, and K.~K. Leung, ``Distributed beamforming and power
  allocation for cooperative networks,'' \emph{IEEE Transactions on Wireless
  Communications}, vol.~7, no.~5, pp. 1817--1822, 2008.

\bibitem{goel2008guaranteeing}
S.~Goel and R.~Negi, ``Guaranteeing secrecy using artificial noise,''
  \emph{IEEE Transactions on Wireless Communications}, vol.~7, no.~6, pp.
  2180--2189, 2008.

\bibitem{Justin}
J.~{Kong}, F.~T. {Dagefu}, and B.~M. {Sadler}, ``Distributed beamforming in the
  presence of adversaries,'' \emph{IEEE Transactions on Vehicular Technology},
  vol.~69, no.~9, pp. 9682--9696, 2020.

\bibitem{liao2010qos}
W.-C. Liao, T.-H. Chang, W.-K. Ma, and C.-Y. Chi, ``{QoS}-based transmit
  beamforming in the presence of eavesdroppers: An optimized
  artificial-noise-aided approach,'' \emph{IEEE Transactions on Signal
  Processing}, vol.~59, no.~3, pp. 1202--1216, 2010.

\bibitem{gershman2010convex}
A.~B. Gershman, N.~D. Sidiropoulos, S.~Shahbazpanahi, M.~Bengtsson, and
  B.~Ottersten, ``Convex optimization-based beamforming,'' \emph{IEEE Signal
  Processing Magazine}, vol.~27, no.~3, pp. 62--75, 2010.

\bibitem{lorenz2005robust}
R.~G. Lorenz and S.~P. Boyd, ``Robust minimum variance beamforming,''
  \emph{IEEE Transactions on Signal Processing}, vol.~53, no.~5, pp.
  1684--1696, 2005.

\bibitem{campi2009scenario}
M.~C. Campi, S.~Garatti, and M.~Prandini, ``The scenario approach for systems
  and control design,'' \emph{Annual Reviews in Control}, vol.~33, no.~2, pp.
  149--157, 2009.

\bibitem{cover1999elements}
T.~M. Cover, \emph{Elements of information theory}.\hskip 1em plus 0.5em minus
  0.4em\relax John Wiley \& Sons, 1999.

\bibitem{choi2017low}
J.~Choi, F.~T. Dagefu, B.~M. Sadler, and K.~Sarabandi, ``Low-power low-{VHF}
  ad-hoc networking in complex environments,'' \emph{IEEE Access}, vol.~5, pp.
  24\,120--24\,127, 2017.

\bibitem{lopez2007semi}
M.~L{\'o}pez and G.~Still, ``Semi-infinite programming,'' \emph{European
  Journal of Operational Research}, vol. 180, no.~2, pp. 491--518, 2007.

\bibitem{gurobi}
\BIBentryALTinterwordspacing
L.~Gurobi~Optimization, ``Gurobi optimizer reference manual,'' 2020. [Online].
  Available: \url{http://www.gurobi.com}
\BIBentrySTDinterwordspacing

\bibitem{sedumi}
J.~F. Sturm, ``Using {SeDuMi} 1.02, a {MATLAB} toolbox for optimization over
  symmetric cones,'' \emph{Optimization Methods and Software}, vol.~11, no.
  1-4, pp. 625--653, 1999.

\end{thebibliography}
\end{document}